\documentclass[11pt]{article}

\usepackage{url}  
\usepackage{graphicx}  
\usepackage{booktabs} 
\usepackage{luca} 
\usepackage{graphicx}
\usepackage{amsfonts,amsmath,amssymb}
\usepackage{hyperref}
\usepackage{comment} 
\usepackage{float} 
\usepackage{tablefootnote} 
\usepackage{xcolor}
\usepackage{algorithm}
\usepackage[noend]{algpseudocode}
\usepackage{geometry} 

\def\mybar#1{\kern-3mm\color{red}\rule{#1cm}{8pt}\kern-#1cm\kern+3mm\color{black}}

\begin{document}

\title{Identifying Fake News from Twitter Sharing Data: A Large-Scale Study}

\author{Rakshit Agrawal$^1$ \and Luca de Alfaro$^1$ \and Gabriele Ballarin$^2$ \and  Stefano Moret$^3$ \and Massimo Di Pierro$^4$ \and  Eugenio Tacchini$^5$ \and Marco L. Della Vedova$^6$\thanks{The authors are listed in alphabetical order.}
\\[1ex]
	\normalsize $^1$ University of California, Santa Cruz, Santa Cruz, CA 95064 USA\\
	\normalsize $^2$ Independent researcher \\
	\normalsize $^3$ \'Ecole Polytechnique F\'ed\'erale de Lausanne, Switzerland \\	
	\normalsize $^4$ School of Computing, DePaul University, Chicago, IL, USA \\
	\normalsize $^5$ Universit\`a Cattolica, Piacenza, Italy. \\
	\normalsize $^6$ Universit\`a Cattolica, Brescia, Italy \\[1ex]
	\normalsize ragrawa1@ucsc.edu, luca@ucsc.edu, gabriele.ballarin@gmail.com, moret.stefano@gmail.com,\\ 
	\normalsize massimo.dipierro@gmail.com,  eugenio.tacchini@unicatt.it, marco.dellavedova@unicatt.it\\
}

\date{February 9, 2019}

\maketitle

\begin{abstract}
Social networks offer a ready channel for fake and misleading news to spread and exert influence.
This paper examines the performance of different reputation algorithms when applied to 
a large and statistically significant portion of the news that are spread via Twitter. 
Our main result is that simple crowdsourcing-based algorithms are able to identify a large portion of fake or misleading news, while incurring only very low false positive rates for mainstream websites.
We believe that these algorithms can be used as the basis of practical, large-scale systems for indicating to consumers which news sites deserve careful scrutiny and skepticism.
\end{abstract}


\section{Introduction}

Social networks and the web have acted as large enablers of open communication in the society.
These platforms have allowed millions of voices to be heard and have further created a network of information dissemination between the people.
While such platforms like Facebook and Twitter have the potential of starting democratic revolutions, when used incorrectly, they also offer misinformation the means to spread, and the resonance chambers where users can consume and reshare them \cite{allcott_social_2017}.
The term \emph{fake news} has emerged distinctively over the past few years, as the spread of targeted and artificially crafted news has plagued the fundamental information that people consume regularly.
This situation has prompted the research community to create the means to identify and label fake news as they spread across social networks; \cite{shu_fake_2017} provides an excellent overview of such research. 

Content-wise, the difference between true and made-up news can be subtle; not infrequently, only the knowledge of details of the situations and people, and only cross-checking sources, enables to tell them apart. 
Sites like Snopes are devoted precisely to this kind of in-depth analysis.
To classify automatically the broad range of news circulated daily on the internet and social media, it can be simpler at least in first approximation to rely on social signals: on the reputation of the sites who published the news, and on the identities of the people who spread them on social media \cite{ma_detect_2015,shao_hoaxy:_2016,tacchini_like_2017,ruchansky_csi:_2017}. 

In this paper, we consider two algorithms based on user and publisher reputation that have been shown to be effective in identifying fake news among a circumscribed set of Facebook posts, and we investigate their effectiveness when applied to a very broad set of news being circulated on Twitter. 
The chief interest of the study lies in its scale. 
Commonly, results on algorithm effectiveness are reported on carefully constructed, and by necessity limited, datasets. 
Even when results on such smaller datasets are promising, the question  remains of showing that what worked in the small, and the models trained from limited ground truth, can then work in the large, when applied to the full spectrum of news being disseminated every day. 
The algorithms we consider, from \cite{tacchini_like_2017}, were shown to work when applied to the set of posts appearing on 32 Facebook pages: 14 hoax-filled pages, and 18 pages devoted to scientific topics. 
Here, we show that they work when applied to a major subset of the news being circulated in English on Twitter. 
The algorithms lead to only a marginal mis-classification of mainstream news (typically 1\% or below), while enabling the identification of a large portion of fake news.
Further, we propose and characterize adaptations of the algorithms that make them practical for real-time classification of such large amount of news.
Thus, the algorithms can provide a practical basis for the large-scale identification of fake news.  

To study the behavior of news classification algorithms in the large, we collected a large subset of the news URLs that were shared on Twitter for several months. 
In particular we collected:
\begin{itemize} 
\item all the news from hundreds of sites that have appeared in published curated lists of sites with dubious reputation or low editorial controls;
\item all the news from a long list of mainstream news sites, including the New York Times, to the Los Angeles Times, to Breitbart, USA Today, Washington Post, and more;
\item all news being spread by a selected group of Twitter users we chose to follow.
\end{itemize}
This initial set was then augmented with the result of queries aimed at dataset completion, finding additional users who shared the news, and additional news shared by users in our dataset.  
The result is a dataset consisting of 5.5 million news articles, 88 million tweets, and 9 million users.
For each news article, we collected both the item title and description, and the identities of the users who shared the item. 
Our experiments are carried out over temporal slices generally consisting of 1.4 million news and 20 to 80 million tweets.

We experimented with two algorithms for identifying fake or misleading news (which we will often simply refer to as ``fake'' from now on).
The first algorithm, from \cite{tacchini_like_2017}, uses as features the identities of Twitter users that spread each piece of news, and optionally words from the title and description of the news item; it then uses a logistic-regression classifier to learn from those features. 
The second algorithm is a graph-based crowdsourcing algorithm from \cite{de_alfaro_reliable_2015,tacchini_like_2017}. 
The second algorithm, termed the {\em harmonic algorithm,} is derived from a {\em boolean crowdsourcing algorithm} that was proposed to aggregate user's True/False (or Yes/No, or Good/Bad) feedback about items into labels for the items \cite{de_alfaro_reliable_2015}. 
The algorithm considers the graph of votes from users to items, and aggregates the user's votes by computing, for each user, a reliability, and for each item, a likely label; indeed, the algorithm bears resemblance to maximum-likelihood approaches.
The algorithm has a computationally efficient structure that makes it applicable to very large graphs, and is able to label large graphs starting from small sets of fake/non-fake seed labels. 
The algorithm was applied to fake news identification in \cite{tacchini_like_2017}, by equating the Yes/True votes to the acts of publishing or sharing the news.

To the best of our knowledge, this paper constitutes the first baseline study of how a fake news detection method fares when applied to the full breath of news being shared on Twitter in a period spanning several months. 
To seed the classification process, we rely on two independently developed lists of low quality news sites: the Opensources list \cite{opensources_curated_2017}, and a list from Metacert\footnote{http://www.metacert.com}; each list comprises about 500 sites. 
These lists are far from exhaustive, and even their union comprises only a fraction of the fake news that circulate every day on Twitter. 
To show that the algorithms are valuable in identifying fake news, we perform three main measures. 

First, as a rough check, we show that if we assume that all news from sites in the seed lists by Opensources and Metacert are fake, and all news from other sites are not, then the harmonic algorithm can classify about 90\% of the news correctly, with the logistic-regression based algorithm offering more unbalanced performance (see Table~\ref{table-res-summary}).
This is an imprecise validation, because the majority of fake news comes from sites not in the Opensources and Metacert lists. 
However, this analysis is still of interest, because statistically most of the news (by share) not appearing in the seed lists come from mainstream news sites (Associated Press, Reuters, mainstream newspapers) that are unlikely to publish fake news.
Classifying 90\% of general news as non-fake provides a basic sanity check.

As further and more insightful validation, we show that the algorithms mis-classify only a very small portion of news from mainstream sites as fake.
The precise amounts depend on the algorithm, and on the list of fake sites we use to seed the algorithms (see Table~\ref{table-news-sites}), but is generally on the order of 1\% at most. 
This is a result of practical significance: too many mis-classifications of news from mainstream sites would quickly erode trust in the classifier predictions. 

As a third measure of the effectiveness of the algorithms, we show that starting from one of the two lists of fake news sites as seed for the algorithms, the harmonic algorithm classifies as fake a majority of the news that are in the other list only. 
This shows that, on the basis of a limited set of seed sites carrying fake news, the algorithms can correctly generalize, and classify as fake a majority of the news that we know via other sources to be fake (see Table~\ref{table-cross-datasets}). 

Finally, we describe an online version of the harmonic  algorithm.
The original algorithm relies on a global fixpoint computation that can be carried out only in batch mode \cite{de_alfaro_reliable_2015}. 
We developed an online version of the algorithm that is able to incorporate information from new tweets and user votes in real time and with minimal computational cost. 
The on-line algorithm propagates updates locally in the graph of users and tweets; we show that this local propagation produces results that are numerically close to those obtained by the fixpoint computation.
This allows us to reduce the frequency of the batch fixpoint computation to few calls or even a single call per day.
Beyond news classification, the on-line version of the algorithm can be used for general boolean crowdsourcing problems, enabling the real-time use of crowdsourced information.

In the interest of reproducibility, and to further facilitate research on the important topic of fake news, we provide all data we used for this paper at \url{https://drive.google.com/open?id=1ziYVTVi5CShCazJzRpHwxYQNO48SogB6}. 

\section{Related Work}

Fake news is a phenomenon that has received wide attention lately, and has motivated much research aimed at detecting fake news automatically; a comprehensive and recent survey is given in \cite{shu_fake_2017}. 
Many approaches are based on the content of the news.
Some of these approaches find their roots in the much older work aimed at detecting spam; see for instance \cite{mason_filtering_2002,vukovic_intelligent_2009}. 
\cite{wang_liar_2017} uses deep learning, LSTMs, and SVMs to classify about 12,000 sort text statements from Politifact; \cite{riedel_simple_2017,ahmed_detection_2017} used a text TF-IDF model followed by a neural net to classify a few thousands text items. 
The writing style has been used in \cite{potthast_stylometric_2017} to classify 1,600 items from BuzzFeed. 

An extensive study was performed in \cite{castillo_information_2011}, which collected tweets verging on 2,500 topics that were trending on Twitter between April and July 2010.  
The topics were classified on the basis of the features of the users and the tweets belonging to the topics, showing high classification accuracy for the topics; the best-performing classification algorithm was based on a decision-tree classifier. 
The social dynamics on Twitter and the temporal propagation of posts has been the subject of a very extensive study in \cite{shao_spread_2017} with the aim of quantifying the effect of bot accounts, detecting them, and tracing how the information spreads. 
The study is similar to ours in extent, involving millions of news items and Twitter users.

The identification of fake news in terms of the reputation both of news publishers, and news sharers, has been suggested in \cite{shu_exploiting_2017}, where the publisher-sharer-news relationship is termed a {\em tri-relationship\/}, and shown to work well on a ground truth of BuzzFeed and PolitiFact news. 
We follow this valuable idea also in the present study, as the harmonic algorithm takes as input both share, and publish, relationship. 

Temporal aspects of news spreading have been studied and correlated to the reliability of news in \cite{ma_detect_2015}, where about a million tweets are used to analyze a hundred news items, and in \cite{shao_hoaxy:_2016}, where large-scale measurements are provided. 
The dynamics of mis-information spread were studied in \cite{del_vicario_spreading_2016}, where the association between groups of users and fake news was noted; this association is at the basis of the algorithms in this paper.

This work is based on the algorithms in \cite{tacchini_like_2017}, which classify news on the basis of the users who shared them. 
The idea of using users as features that drive the classification has also been used in \cite{ruchansky_csi:_2017}, where user identities are fed along with text into a deep-learning system.

\section{Dataset}

To characterize the performance of fake news detection algorithms for news spread via Twitter, it is essential to have a dataset that covers the full spectrum of news that are shared on Twitter, including both mainstream sites and sites whose quality is less certain.  
Since our concern is news classification, we restricted our attention to tweets that contain at least one URL. 
Ideally, we would like our dataset to include all URL-containing tweets;  however, acquiring this much data would be prohibitively expensive.
We therefore settled on a sampling strategy that allowed us to collect a large number of news-related tweets and re-tweets, and build a significative sub-graph.
We achieved this by gathering URL-containing tweets using the Twitter streaming API. We collected all Tweets from August 1, 2017 until December 8, 2017 that satisfied at least one of the following filters:
\begin{itemize}

\item {\em Mainstream news:} all tweets containing URLs from a list of mainstream news sites, including ABC News, Breitbart, BuzzFeed, CBS News, Channel 7 News, CNN, Fox News,  MSNBC, NBC News, The Huffington Post, The Economist, The Guardian, The Hill, The Onion, The New York Post, The New York Times, The Times, The US Herald, The Washington Post, USA Today, and US News;
 
\item {\em News agencies:} all tweets from The Associated Press and Reuters;
 
\item {\em Scientific, and peer-reviewed news:} all tweets from ArXiV, Nature, and Science Magazine;

\item {\em Fact-checking sites:} all tweets from Politifact and Snopes;
 
\item {\em News from selected users:} all tweets from a set of about 160 users, which included many of the most active tweeters in news and science; and 
 
\item {\em Low-quality news:} all tweets that mentioned URLs from many of the sites mentioned as fake, bias, unreliable, clickbait, or consipiracy in the lists \cite{zimdars_false_2016,opensources_curated_2017}.  The original lists consisted of 1,000 sites; we excluded many that seemed to be no longer active, or that published news only sporadically.

\end{itemize}
In addition, we ran specific queries aimed at completing the graph, by periodically querying (within the limits of the Twitter API) for URLs that were mentioned only once or twice, in the attempt to find more tweets referencing these URLs; and users from whom we had a limited number of tweets, in the attempt to discover further activity from such users. 
Overall, between the streaming and the graph-completion queries, we gathered about a million URL-containing tweets per day. 
Upon gathering each tweet, we parsed it, and we downloaded the HTML of the reference page.
From the semantic tags ({\tt og:url}, {\tt og:title}, and {\tt og:description}) in the HTML we extracted and stored the canonical URL, the title, and the description. 

We report in Table~\ref{table-dataset-composition} the list of news sites with the most URLs in the dataset for the period from September 1 to November 30, 2017.

\begin{table}
\centering
\begin{tabular}{l|r||l|r}
Site & \% & Site & \% \\ \hline
                   youtube.com  & 4.319 &                  wordpress.com  & 0.754  \\
                   nytimes.com  & 3.145 &                     nypost.com  & 0.625  \\
               theguardian.com  & 2.904 &                    thehill.com  & 0.619  \\
            huffingtonpost.com  & 1.964 &                    latimes.com  & 0.616  \\
            washingtonpost.com  & 1.944 &                  breitbart.com  & 0.609  \\
                     arxiv.org  & 1.585 &                    cbsnews.com  & 0.563  \\
                  usatoday.com  & 1.504 &                    reuters.com  & 0.426  \\
                indiatimes.com  & 1.458 &                     reddit.com  & 0.388  \\
                   foxnews.com  & 1.262 &                dailycaller.com  & 0.367  \\
                  blogspot.com  & 1.202 &                    newsmax.com  & 0.336  \\
\end{tabular}

\caption{The 20 news sites with the most URLs in our dataset in the period from September 1 to November 30, 2017. }
\label{table-dataset-composition}
\end{table}


Once we obtained the tweets and the corresponding URLs, we processed them, and we associated each URL with the following information:
\begin{itemize}
\item {\em URL date:\/} the date at which we first saw the URL. 
\item {\em Users:\/} the list of Twitter usernames who shared the URL. 
\item {\em Title:\/} the title of the news article, as given in the {\tt og:title} social tag.
\item {\em Description:\/} the description of the news article, as given in the {\tt og:description} tag.
\end{itemize}
The data used for the paper can be found at\\ \url{https://drive.google.com/open?id=1ziYVTVi5CShCazJzRpHwxYQNO48SogB6}. 

\subsection{Seed Fake News Sites}

All the algorithms that we utilize fall into the category of semi-supervised learning: in order to train them we need a set of example URLs that we can reliably categorize as good (reliable news) or bad (fake or misleading news).  
Ideally we would like to utilize a large, unbiased, and possibly crowdsourced, ground truth.
Unfortunately, such a ground truth is not currently available to the authors. 
Most sets of classified URLs for fake news are fairly small, generally numbering less than 1,000 URLs.
It is difficult to build, on the basis of such small sets, algorithms that generalize well to the whole set of news being shared on Twitter.
Furthermore, and perhaps even more importantly, if the testing set is small it does not yield useful information on the algorithms' performance because we cannot extrapolate the results to the full set of URLs collected on Twitter, which is the question that interests us here.
 
As the next-best solution, we decided to use for our algorithms sets of seed fake labels consisting of two curated lists of news sites which comprise of (in the opinions of the curators) fake, misleading, unchecked, and biased news: the {\em Metacert\/} list, and the {\em Opensources\/} list.
\begin{itemize}

\item{\em Opensources list.}
The {\em Opensources\/} list comprises of sites derived from a list originally posted by Melissa Zimdars \cite{opensources_curated_2017}.
We kept all sites in that list with type {\em bias, fake, junksci, hate, conspiracy,} or {\em clickbait.}
This resulted in 581 sites.

\item{\em Metacert list.}
The {\em Metacert\/} list comprises of sites flagged as fake, misleading, or unreliable news sources using a crowdsourcing method by Metacert\footnote{https://metacert.com/}, and kindly shared with the authors. 
The list consists of 500 sites.

\end{itemize}
These two lists have partial overlap, sharing 331 sites. 
The algorithms we consider will be trained on these fake seed lists as fake labels, and on all other URLs (or a random sample of them) as non-fake labels, as will be described in detail in the following section. 

Using these two lists as seed fake labels entails two problems. 
The first problem is that the lists consist of sites, not of URLs. 
There is thus the risk that our algorithms simply learn the sites, rather than the characteristics of fake news. 
We have taken two steps to counter this. 
As mentioned above, we ensure that no information used in the algorithms directly mentions the site. 
In particular, we remove all mentions of the sites where the news appear (and their aliases) from the content we use as algorithm input.

The second problem is that the Metacert and Opensources lists are only partial lists of sites containing fake or misleading information. 
This is why we call them {\em seed\/} lists: they are a good starting point, but far from complete.
Even when we use the union of the two lists, we cannot assume that the complementary set of URLs contain exclusively reliable news. 
We counter this problem in three ways. 
First, we see from Table~\ref{table-dataset-composition} that a majority of URLs in our dataset comes from mainstream sites that are unlikely to publish fake news. 
Furthermore, and more importantly, we will present not only aggregate precision and recall results, but also detailed results for many individual web sites (see Table~\ref{table-news-sites}).
This will make it clear how, for mainstream journalistic sites, the percentage of news mis-classified as fake is quite low. 
Lastly, we will show results on {\em cross-learning,} which show that given one of the seed lists, the algorithms can recover as fake a majority of the URLs that belong to the other seed list {\em only.}
In other words, from one fake seed list, we can recover a majority of the sites and URLs present in the other. 
Together, these results will show how the algorithms are good at detecting fake news, while sparing mainstream media from mis-identifications, even when presented with a very broad set of new shared on Twitter. 

\begin{figure}
	\begin{center}
		
\includegraphics[width=0.55\textwidth]{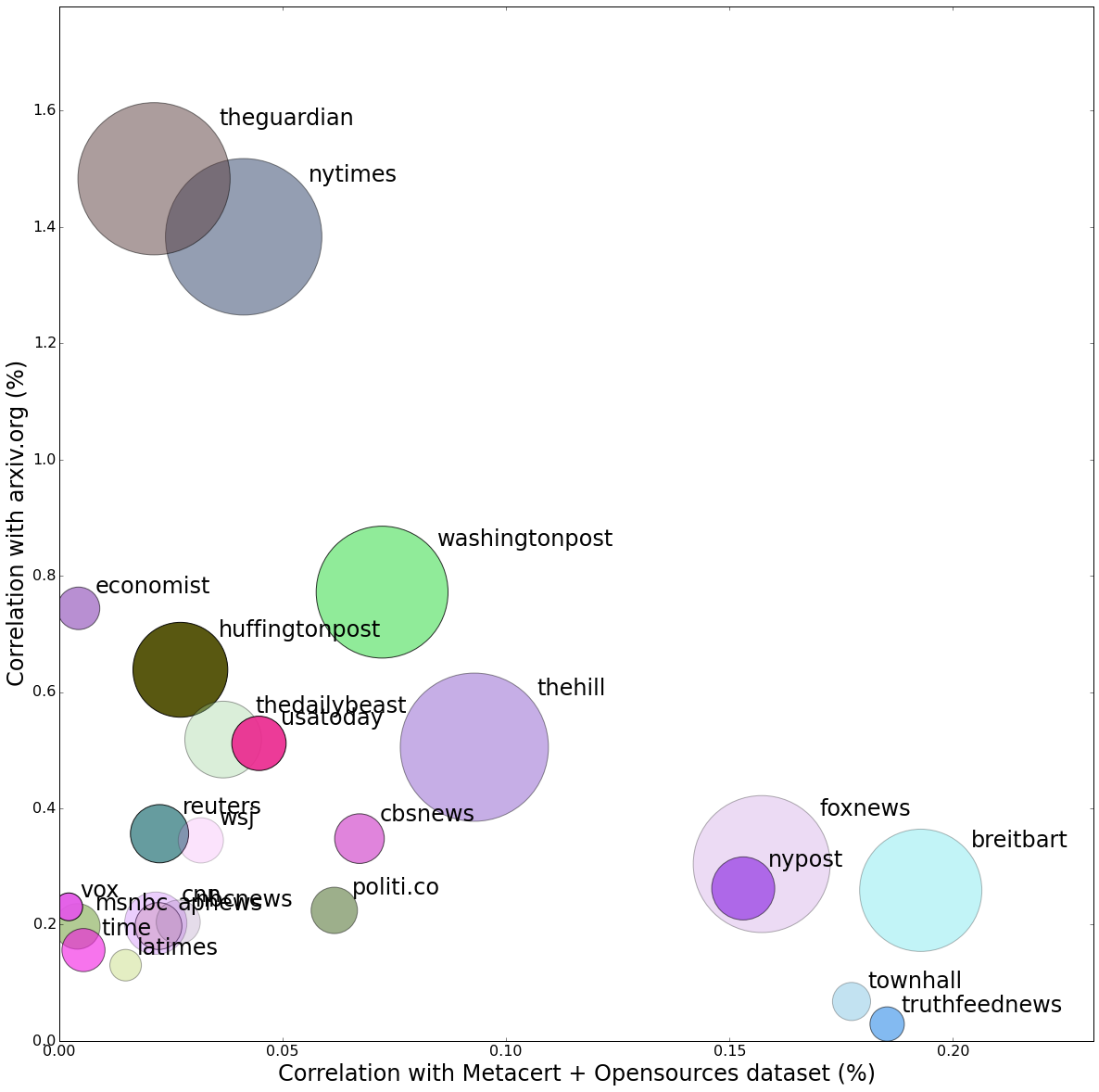}

\caption{Average correlation of most tweeted news sites with known fake news sites (Opensources + Metacert lists) vs correlation with science related (arxiv.org).  The area of each circle is proportional to the number of URLs we have from each site. The data is from November 1 to November 7, 2017.}
\end{center}
\label{fig-circles}
\end{figure}

To illustrate the range of news collected, we give in Figure~\ref{fig-circles} the correlation between the main news sites, and the sites appearing in the union of the Opensources and Metacert lists on one side, and the scientific site arxiv.org on the other. 
We compute the correlation between two sites $i,j$ as $T_{ij}/\sqrt{T_i T_j}$ where $T_{ij}$ counts the number of users that tweeted about both domains weighted by their number of tweets; $T_i$ and $T_j$ count the number of users that twitted about domain $i$ and $j$ respectively, also weighted by the number of tweets. 

\section{Methods}

We experimented with two different approaches in building a reputation system for news shared on Twitter 
In the first approach, we considered the identity of each user who shared the news article, and each word that appears in the title or description of the article itself, as an individual feature.  
This gives rise to a large number of features which we used to train a logistic regression classifier \cite{tacchini_like_2017}. 
In the second approach, we implement the ``harmonic'' crowdsourcing algorithm described in \cite{de_alfaro_reliable_2015,tacchini_like_2017}.
The latter approach can also dynamically accommodate explicit votes expressed by users on the news items. 

\subsection{Logistic-regression classifier}
\label{sec-logistic}

\subsubsection{Architecture.} 

Our first and simplest model is from \cite{tacchini_like_2017}: we use as list of features for a piece of news (a URL after canonicalization) the individual users who shared the URL by tweeting or retweeting it, alongside with the words that appear in the title and description of the news article itself. 
This gives rise to a large set of features, numbering in the hundred of thousands: each user and each word is a feature. 

On this large, and rather sparse feature set, we train a simple logistic regression model. 
Logistic regression models are well suited to handle large and sparse feature sets. 
Intuitively, the coefficient associated with each user is positive if the user tends to share high-reputation news, and negative if the user tends to share low-reputation news. 
Similarly, the coefficient of a word is positive if the word tends to appear in the title or description of high-reputation news, and negative if the word tends to appear in the title or description of low-reputation news. 

We consider three kind of models: 
\begin{itemize}
\item {\bf LR-U}, built on users' identities only.
\item {\bf LR-UT}, built on users' identities and words in title/description.
\item {\bf LR-T}, built on words in title/description only.
\end{itemize}
Models are implemented using the {\tt scikit-learn\/} Python package. 
Since our labels are based on sites, rather than individual URLs, we removed from the title and descriptions all mentions of the news site to which the individual URLs belong, to avoid the corresponding bias in our experiments.

\subsubsection{Training and testing datasets.}

To evaluate the logistic-regression based classification methods, we split the dataset into {\em training\/} and {\em testing\/} datasets, as follows.
\begin{itemize}
\item The full training dataset consists of all URLs that have been first seen from September 1st, 2017, to October 20, 2017. 
For each URL, we keep all the tweets whose timestamp is strictly before October 30, 2017.

\item The full testing dataset consists of all URLs that have been first seen from October 31, 2017, to November 26, 2017. 
For each URL, we keep all the tweets that mention it.
\end{itemize}
The construction of the full training and testing datasets ensures that the URLs they contain are disjoint, and that the testing dataset is built only from information that chronologically precedes all of the information in the testing dataset. 

The interval from October 20 to October 30, 2017, allows us to collect tweets mentioning the URLs seen first on October 20 for 9 additional days, until midnight of October 29 (all dates used in the experiment are in UTC). 
As the full training and testing datasets are quite large, we subsampled them by keeping only the tweets appearing on alternating days, starting from September 1, and October 31, respectively. 

The characteristics of the resulting datasets are reported in Table~\ref{table-reduced-datasets}.
The composition of the two lists of labels is summarized in Table~\ref{table-gt-sizes}.
As the number of positive and negative instances in the training set is highly unbalanced (in the training dataset, only 21,263 of 787,601 URLs are considered fake news), we train the logistic classifier giving class weights that are inversely proportional to class sizes, thus minimizing per-class errors. 

\begin{table} 
\centering
\begin{tabular}{l|r|r}
Dataset & URLs & Tweets \\ \hline
Training dataset & 787,601 & 14,587,984 \\
Testing dataset & 607,299 & 7,967,170
\end{tabular}

\medskip
\caption{Sizes of training and testing datasets for logistic-regression based classifiers.}
\label{table-reduced-datasets}
\end{table}

\begin{table}
\centering
\begin{tabular}{l|r|r|r|r}
& \multicolumn{4}{|c}{Number of URLs} \\
& Opensources & Metacert & Common & Total \\ \hline
Train & 21,263 & 21,146 & 14,339 & 787,601 \\
Test  & 13,388 & 11,740 & 9,210 & 607,299 \\
\end{tabular}

\medskip
\caption{Number of URLs in the Opensources and Metacert  lists that appear in our training and testing sets for logistic-regression classifiers, alongside the total number of URLs in the datasets.}
\label{table-gt-sizes}
\end{table}

\subsection{Crowdsourcing-based classifier}

The second classification method we used is based on the crowdsourcing-inspired ``harmonic boolean label crowdsourcing'' model of \cite{de_alfaro_reliable_2015,tacchini_like_2017}. 
In boolean label crowdsourcing models, users provide True/False labels for a set of items, and the crowdsourcing algorithms then attribute to each item an estimate of truth or falsehood, and to each user an estimate of how much the user is likely to tell the truth or to lie \cite{karger_iterative_2011,liu_variational_2012}.
As in \cite{tacchini_like_2017}, we use a variant of those methods in which the truth or falsehood of a subset of items is taken as known; the algorithms then propagate the information to the remaining users and items. 
We use the {\em harmonic} algorithm presented in \cite{de_alfaro_reliable_2015}, both because it is efficient on very large-scale data, and because in \cite{tacchini_like_2017} it has been shown to give good results.

\subsubsection{The harmonic algorithm.}

We represent our entire dataset as a bipartite graph $(I \union U, T)$, where $I$ are the news items, $U$ are the users, and $T \subs I \times U$ are the edges. 
With each edge $(i, u) \in T$ is associated a {\em polarity\/} $p_{i,u} \in \set{-1,1}$. 
Edges arise in the graph in three ways: tweets, editorial, and vote. 
\begin{itemize}
\item {\em Tweet:\/} A tweet by user $u$ containing the URL of item $i$; we have $p_{i,u} = 1$.
\item {\em Editorial:\/} For each news item $i$ whose URL belongs to a site $d$, we add an edge $(i,d)$ to the graph with $p_{i, d} = 1$. 
The edge represents the editorial approval of $d$ (for instance, \url{www.nytimes.com}) for the news item $i$ it published. 
We create such editorial edges for all sites, except those that consist in content aggregators without editorial oversight (such as \url{www.youtube.com}, \url{www.instagram.com}, for instance).
\item {\em Vote:\/} a vote by user $u$ on the truth of $i$.  In this case, $p_{i,u} = 1$ if $u$ votes $i$ truthful, and $p_{i,u} = -1$ if $u$ votes $i$ as fake.
\end{itemize}
Our use of the editorial relation between a site and the news it publishes is similar to the tri-relationships of \cite{shu_exploiting_2017}.
However, when reporting the accuracy of our classifier, {\em we will exclude editorial edges from the model.} 
Since our labels are based on sites, connecting all news items from a site to a node representing the site via editorial edges would make the classification problem improperly easier.
If we included editorial edges, we would be able to classify correctly both fake and non-fake with accuracy above 95\%, rather than the approximately 90\% we obtain without editorial edges.
Likewise, we omit vote edges to make the results more comparable with those obtained via logistic regression.

We denote by $\partial i = \set{u \mid (i, u) \in L}$ and $\partial u = \set{i \mid (i, u) \in L}$ the 1-neighborhoods of an item $i \in I$ and user $u \in U$, respectively.
The harmonic algorithm classifies all the items by running a global fixpoint over the bipartite graph of items and users. 
For each node $v \in I \union U$, the algorithm iteratively computes two non-negative parameters $\alpha_v$, $\beta_v$ defining a beta distribution: intuitively, for a user $u$, $\alpha_u - 1$ represents the number of times we have seen the user share a reliable news item, and $\beta_u - 1$ represents the number of times we have seen the user share a fake or misleading news item.
Similarly, for a news item $i$, $\alpha_i - 1$ is the number of shares from reliable users, and $\beta_i - 1$ the number of shares from unreliable users. 
For each node $v$, let $q_v = (\alpha_v - \beta_v) / (\alpha_v + \beta_v)$: for a news item $i$, positive values of $q_i$ indicate a likely reliable news, and negative values, a likely fake. 

The training set consists of two subsets $I_F, I_N \subs I$ of fake and non-fake news. 
Initially, the algorithm sets $q_i := -1$  for all $i \in I_F$, and $q_i := 1$ for all $i \in I_N$; it sets $q_i = 0$ for all other posts $i \in I \setm (I_F \union I_N)$.
The algorithm then proceeds to iteratively propagate the information from items to users, and from users to items. 
First, for each user $u \in U$, it lets: 
\begin{align*} 
    \alpha_u & := c + \sum \set{p_{i,u} q_i \mid i \in \partial u, p_{i,u} q_i > 0} \\
    \beta_u & :=  c - \sum \set{p_{i,u} q_i \mid i \in \partial u, p_{i,u} q_i < 0} \eqpun . 
\end{align*}
The positive constant $c = 0.02$ acts as regularization. 
In the second part of the iteration, the algorithm updates the values for each item $i \in I \setm (I_F \union I_N)$ (without affecting the seed labels) by:
\begin{align*} 
    \alpha_i & := c + \sum \set{p_{i,u} q_u \mid u \in \partial i, p_{i,u} q_u > 0} \\
    \beta_i & :=  c - \sum \set{p_{i,u} q_u \mid u \in \partial i, p_{i,u} q_u < 0} \eqpun . 
\end{align*}
We experimented with various numbers of iterations, and we settled on using three (full) iterations, which seem sufficient to spread information from the labeled nodes.
Once the iterations are concluded, we classify a news item $i$ as fake or misleading if $q_i < 0$, and as high reputation (or reliable) otherwise; thus, we take $q_i$ as the reputation of $i$.

\subsubsection{Training and testing datasets.}

We built the bipartite graph for the harmonic algorithm using the 61,204,978 tweets collected between September 1, 2017, and December 8, 2017, the same period used for the logistic approach. 
The training set consists of the 1,814,741 URLs which occurred in tweets from September 1 to October 20; the testing set consists of the 1,192,934 URLs whose first occurrence was recorded from November 1 to November 26, included.
Again, these dates are the same as those used in testing the logistic classifier. 
For the harmonic algorithm, due to its greater scaling ability, we were able to run it on the complete dataset, rather than having to restrict it to the URLs occurring on alternate days.\footnote{The URLs are not exactly in a factor-of-two ratio with those of the logistic classifier, as we occasionally had days where the system was down for maintenance.}

In the training set, we take as negative labels all URLs that belong to the Opensources or Metacert lists; let $n$ be the number of such URLs. 
As $n \ll 1,814,741$, if we labeled positively all the remaining URLs in the training set, the training set would have a strong class imbalance.
Rather, we selected randomly $\kappa n$ URLs non in these lists and take them as positive labels. 
The multiplier $\kappa$ enables us to control the bias of the classifier; we report the results for $\kappa=1$ and $\kappa=2$.

\subsubsection{Online classification.}

The logistic classifier is well suited for classifying news items as more information (more tweets) is received.
To do so, for each item $i$ the classifier stores the total $s_i = w_0 + \sum_{u \in \partial i} w_u$, where $w_u$ is the weight of user $u$ in the logistic classifier and $w_0$ is the bias of the logistic regression.
When we receive a new tweet by user $u$ about item $i$, we update $s_i := s_i + w_u$, and we classify $i$ according to the sign of $s_i$. 

In contrast, the harmonic classifier operates by computing a global fixpoint, which is a time-consuming operation. 
In our system, computing the fixpoint takes about one hour, and the fixpoint is computed three times a day.
Between these computations, the harmonic classifier is unable to process information in real time to improve news classification.
To overcome this limitation, we present here an on-line adaptation of the algorithm, based on the idea of propagating local updates to the graph as edges are added.
The online version of the harmonic algorithm stores the $q, \alpha, \beta$-values for all items and all users. 
%
%
When an edge $(i, u)$ is added to the graph due to a tweet or vote, we propagate the net contribution $q_u p_{i,u}$ of $u$ to $i$ by calling \textsc{UpdateItem}$(i, q_u p_{i,u}, \ell, \kappa)$ in Algorithm~\ref{algo-online}, where the parameter $\ell$ represents the maximum depth of propagation, and $\kappa > 0$ the minimum change that is worth propagating. 
In our live system we use $\ell = 1$ and $\kappa = 0.02$. 

\begin{algorithm}
\caption{Dynamic harmonic propagation}
\label{algo-online}
\begin{algorithmic}[1]
\Procedure{UpdateItem}{$i, \delta, \ell, \kappa$}:
  \State If $\delta > 0$: $\alpha_i := \alpha_i + \delta$
  \State If $\delta < 0$: $\beta_i := \beta_i - \delta$
  \State $q'_i := (\alpha_i - \beta_i) / (\alpha_i + \beta_i)$
  \State $\delta' := q'_i - q_i$
  \State $q_i := q'_i$
  \If {$\ell > 0$ and $|\delta'| \geq \kappa$}
    \For{$u \in \partial i$}
      \State UpdateUser$(u, \delta', \ell - 1, \kappa)$
    \EndFor
  \EndIf
\EndProcedure
\Procedure{UpdateUser}{$u, \delta, \ell, \kappa$}:
  \State If $\delta > 0$: $\alpha_u := \alpha_u + \delta$
  \State If $\delta < 0$: $\beta_u := \beta_u - \delta$
  \State $q'_u := (\alpha_u - \beta_u) / (\alpha_u + \beta_u)$
  \State $\delta' := q'_u - q_u$
  \State $q_u := q'_u$
  \If {$\ell > 0$ and $|\delta'| \geq \kappa$}
    \For{$i \in \partial u$}
      \State UpdateItem$(i, \delta', \ell - 1, \kappa)$
    \EndFor
  \EndIf
\EndProcedure
\end{algorithmic}
\end{algorithm}

In the next section we report results showing that as long as the fixpoint computation is run periodically, the on-line adaptation is capable of updating the classification in real time, processing both tweets and votes as they are received, while inducing only a very small deviation from the results that would have been obtained in the ideal case of running the fixpoint each time new information is received. 
We note that the above on-line adaptation can be used both in news classification, and also to obtain an on-line version of the boolean crowdsourcing algorithms presented in \cite{de_alfaro_reliable_2015}.

\section{Results}
\label{sec-results}

\begin{table}[t]
  \centering
    \begin{tabular}{r||r|r|r||r|r}
      
      Recall & LR-U       & LR-UT     & LR-T      &HM-2x    & HM-3x \\ \hline
      Fake:           & 57.64      & 61.14     & 57.25     & 89.81   & 86.35             \\
      Non-Fake:        & 97.40      & 97.75     & 94.18     & 88.14   & 91.58    
    \end{tabular}

  \medskip
  \caption{Fake and non-fake recall expressed as percentages. LR-U, LR-UT, LR-T are the logistic regressions based on users (-U), text (-T) and users plus text (-UT). HM-2x and HM-3x are the harmonic crowdsourcing algorithm with 2x, and 3x sampling of good URLs. These results are based on the Opensources list. 
  }
  \label{table-res-summary}
\end{table}

Table~\ref{table-res-summary} summarizes the results for the various algorithms under consideration. 
For each algorithm, and each test-set, we report: 
\begin{itemize}

\item {\bf Fake recall:} the percentage of news items in the list of seed fake news sites that the algorithm labels as low reputation.

\item {\bf Non-fake recall:} the percentage of news items that do not belong to the seed lists of fake news that the algorithm labels as high reputation.

\end{itemize}
The fake and non-fake recalls, together, give a picture of the classification for URLs that belong, or not belong, to the seed lists.
Unfortunately, as we remarked in the introduction, the Opensources and Metacert lists are only partial lists of fake news sites.
For this reason, the non-fake recall is only meaningful as a sanity check.
Most of the news in our dataset come from mainstream news sites, and classifying too many of them as fake would be problematic: thus, it is a good sign that the non-fake recall is on the order of 90\%. 
Still, the news items not in the fake seed lists contain many fake news, so that even for a perfect classification, the non-fake recall, defined on the basis of the lists, would not be 100\%.
Further, we cannot give meaningful precision figures, because the two lists of fake sites we use contain only a small subset of the fake news that are spread every day on Twitter.
Indeed, news that are detected as fake but do not belong to the lists are often fake or dubious. 
We return to this point in Section~\ref{sec-crossprediction}, where we show the ability of one list to predict sites belonging exclusively to the other list.
The performance on individual news sites, described below, also offers a more precise picture of detection accuracy.

\subsection{A look at individual news sites}

To gain a better understanding of the performance of the algorithms, it is useful to look at the results for individual news sites not in the fake lists, given in Table~\ref{table-news-sites}.
The data presented in Table~\ref{table-news-sites} gives insight on the key question: what would happen if we were to apply the fake detection algorithms to the full breath of news shared via Twitter? 
While in the table we report results only for a subset of sites, we included major news sites, along with news agencies, and along with several sites of varying degree of editorial controls. 
The information Table~\ref{table-news-sites} provides is thus complementary to that reported in studies that rely on small, curated ground truths. 
Small, curated ground truths give information on precision and recall {\em with respect to the ground truth.}
Data such as the one in Table~\ref{table-news-sites} provides information on the classification {\em with respect to the breath of news being shared every day.}
If one were to turn the algorithms into ``productized'' detection mechanisms for fake news, the table provides insight into what kind of content would be flagged.

For most mainstream news sites, as well as for sites that publish scientific articles such as \url{arxiv.org}, few news are labeled as low reputation.
Further, many sites that have a high percentage of URLs labeled as low reputation are not known for their high standards of journalistic accuracy and editorial controls. 
Indeed, our algorithms seem to be able to use the Opensources and Metacert list to discover other unreliable sites not originally in these lists.
This occurs because the reputation systems assign low reputation (low factors) to the users who share known unreliable news, and then follow those users to other news they have shared.

\begin{table*}[t]
	\footnotesize
\centering
\begin{tabular}{l|r|r||r|r||r|r||r|r||r|r}
  & \multicolumn{2}{c||}{LR-UT} & \multicolumn{2}{c||}{LR-U} & \multicolumn{2}{c|}{HM-2x}& \multicolumn{2}{c|}{HM-3x} & \multicolumn{2}{c}{HM(EE)-2x} \\
  & OS           & MC          & OS          & MC          & OS               & MC              & OS               & MC             & OS          & MC \\ \hline
  nytimes.com               & 0.88         & 0.83        & 0.83        & 0.73        & 3.26             & 3.09         & 2.56         & 1.14       & 0.84       & 0.10\\
  theguardian.com           & 1.16         & 1.02        & 0.97        & 0.93        & 2.74             & 2.59         & 1.86         & 0.99        & 0.53       & 0.05\\
  huffingtonpost.com        & 2.61         & 1.96        & 1.43        & 1.40        & 3.90             & 3.83         & 3.06         & 1.68   & 0.74       & 0.15\\
  washingtonpost.com        & 1.92         & 2.12        & 3.07        & 2.98        & 5.59             & 5.31         & 4.46         & 1.86  & 1.69       & 0.19 \\
  arxiv.org                 & 0.08         & 0.08        & 0.11        & 0.11        & 0.09             & 0.10         & 0.07         & 0.03   & 0.01       & 0.00  \\
  usatoday.com              & 1.09         & 1.13        & 0.84        & 0.63        & 5.21             & 3.91         & 4.30         & 1.44   & 2.24       & 0.15       \\
  foxnews.com               & 1.91         & 2.01        & 3.07        & 2.77        & 41.37            & 37.04         & 30.40         & 7.81   & 32.48       & 2.69  \\
  nypost.com                & 3.42         & 3.20        & 3.70        & 3.36        & 19.91            & 18.35         & 15.37         & 5.86 & 10.60       & 0.93   \\
  thehill.com               & 3.39         & 3.43        & 4.64        & 3.55        & 12.34            & 11.99         & 10.30         & 4.20  & 5.89       & 0.89  \\
  latimes.com               & 0.21         & 0.17        & 0.38        & 0.42        & 2.31             & 2.26         & 1.84         & 0.89  & 0.68       & 0.14    \\
  breitbart.com             & 5.10         & 3.39        & 11.04       & 9.37        & 85.42            & 85.14         & 80.59         & 28.35   & 90.15       & 15.74 \\
  cbsnews.com               & 1.53         & 1.88        & 1.62        & 1.22        & 9.05            & 8.57          & 7.33         & 3.32  & 2.81       & 0.39\\
  reuters.com               & 1.16         & 1.11        & 1.07        & 0.93        & 2.22             & 2.02         & 1.73         & 0.85   & 0.45       & 0.15\\
  dailycaller.com           & 25.00        & 41.38       & 50.58       & 56.59       & 95.82            & 95.46        & 94.20         & 77.66     & 97.25       & 7.34    \\
  townhall.com              & 31.41        & 16.74       & 38.88       & 23.49       & 84.75            & 85.35         & 82.18         & 52.70  & 87.46       & 32.57      \\
  truthfeednews.com         & 21.31        & 2.95        & 18.82       & 1.87        & 99.91            & 99.93         & 99.89         & 99.45    & 99.98       & 99.89 \\
  hotair.com                & 9.40         & 2.03        & 15.60       & 10.15       & 62.53            & 58.30         & 55.40         & 26.80   & 53.41       & 1.72     \\
  freedomdaily.com          & 88.89        & 87.96       & 81.48       & 78.70       & 98.54            & 98.39          & 97.73         & 85.58   & 99.93       & 98.76   \\
  conservativedailypost.com & 93.20        & 92.37       & 84.54       & 81.03       & 99.02            & 98.93         & 98.68         & 91.71   & 99.92       & 99.61 \\
  lucianne.com              & 15.65        & 96.56       & 89.31       & 96.56       & 99.75            & 99.70         & 99.65         & 98.75        & 99.95       & 1.06\\
  redstate.com              & 39.23        & 14.67       & 39.23       & 18.18       & 84.95            & 64.51         & 73.99         & 3.94        & 83.85       & 3.98\\
  theblaze.com              & 43.06        & 24.07       & 16.67       & 12.50       & 82.41            & 80.10          & 74.69         & 36.63         & 75.99       & 9.88 \\
  newsbusters.org           & 18.22        & 21.78       & 29.33       & 31.56       & 96.49            & 96.54         & 95.01         & 62.45         & 98.15       & 4.71 \\
  zerohedge.com             & 28.18        & 19.86       & 31.18       & 23.33       & 74.58            & 72.24         & 65.55         & 29.95       & 63.45       & 8.63
\end{tabular}

\medskip
\caption{
Percentage of URLs that are classified as low reputation in a representative sample of news sites when seeding the classification with the Opensources (OS) and Metacert (MC) lists.
The methods are logistic regression based on users (LR-U), text (LR-UT), and both users and text (LR-UT); and harmonic crowdsourcing based on 2-fold (HM-2x) and 3-fold (HM-3x) sampling of good URLs.
The algorithm HM(EE)-2x indicates harmonic crowdsourcing with 2-fold sampling using the editorial edges: this is the algorithm we use in our live system. 
}
\label{table-news-sites}
\end{table*}

\subsection{Cross-list prediction}
\label{sec-crossprediction}

\begin{table}
\centering
\begin{tabular}{l|l|c||r|r|r}
   &       & N. of   & \multicolumn{3}{c}{Detection} \\
   &       & \multicolumn{1}{c||}{URLs in} & \multicolumn{1}{c|}{Direct} &    \multicolumn{1}{c|}{Suspicious} & \multicolumn{1}{c}{Suspicious} \\
List & Method  & diff    & \multicolumn{1}{c|}{URL}  & \multicolumn{1}{c|}{Site} & \multicolumn{1}{c}{URL} \\ \hline \hline

OS & LR-UT & 2530 & 21.7 & 91.7 & 95.5 \\
OS & LR-U  & 2530 & 42.6 & 75.0 & 93.3 \\
OS & HM-2x  & 4158 & 71.5 & 70.6 & 86.2 \\
MC & LR-UT & 4178 & 15.0 & 76.7 & 88.9 \\
MC & LR-U  & 4178 & 18.3 & 76.7 & 83.9 \\ 
MC & HM-2x  & 16624 & 80.6 & 93.1 & 94.8 \\ 
\end{tabular}
\medskip
\caption{Cross dataset detection percentages of fake news and misleading URLs, starting from the Opensources (OS) and Metacert (MT) lists.}
\label{table-cross-datasets}
\end{table}

To validate the ability of the reputation systems for fake news to discover additional unreliable news and sites, we conducted an experiment.
The two lists of fake sites at our disposal, which have been developed independently, share only 331 of 581 (Opensources) and 500 (Metacert) sites respectively.  
Thus, we ask the question: starting from one list, how good are our reputation systems at discovering the additional URLs and sites in the other list? 
From Table~\ref{table-news-sites}, we see that sites with over 5\% of their URLs classified as fake are of uncertain quality, worthy of manual investigation by a human. 
We measured the following cross-list detections:
\begin{itemize}
\item {\em Direct URL detection:} what percentage of URLs appearing only in the other list do we label as fake? 
\item {\em Suspicious site detection:} among the sites appearing only in the other list and that have at least 20 URLs in the dataset, how many of them do we detect as potentially fake or misleading, by flagging over 5\% of their URLs as fake? 
\item {\em Suspicious URL detection:} what percentage of URLs appearing only in the other list belong to suspicious sites detected as above? 
\end{itemize}
For suspicious site detection, the limitation of sites with at least 20 URLs removes both inactive sites, and sites on which we have so few URLs that applying the 5\%-test is unreliable.
The results are given in Table~\ref{table-cross-datasets}. 
As we see, direct URL detection is low: this is in line with the results in Table~\ref{table-news-sites}, where we see that only part of URLs belonging to other fake or misleading sites are flagged as fake.
On the other hand, the detection of suspicious sites, and suspicious URLs, is high.
We also observe that the Opensources list is more predictive of the Metacert list than the other way round. 
This illustrates how reputation-based methods can be effective in flagging sites for further investigation, on the basis of the news-resharing behavior of users that connects them to sites appearing in list of fake news sites.

\subsection{Online classification}

To verify the accuracy of our online adaptation of the harmonic algorithm, we performed a set of experiments on data containing editorial edges, to reflect our use in the online system.
Using the harmonic algorithm, we first computed a global fixpoint on data for 20 days starting from September 1, 2017.
Then, we added the edges for one day at a time (about 800,000 edges each day), propagating the updates with our online adaptation of the harmonic algorithm. 
We then performed a daily global fixpoint computation, and we measured the difference in valuation of news sites before, and after, such global fixpoint valuation: this gave us a measure of the accuracy attained by the online algorithm, when performing global computation only once a day.
We repeated this for 10 consecutive days.
In Table~\ref{table-online-classification}, we report the number and percentage of items whose online value is within 0.1 of the fixpoint one (in a range from $-1$ to $+1$).
We give the data for all items, for items that are first seen on a given day, and for items that were tweeted on a given day.
We repeated the experiment with online propagation being performed every 2 days, and we observed similar results.
We did not notice appreciable advantages in setting $\ell = 2$ or choosing a value for $\kappa$ smaller than 0.02. 
Together, $\ell=1$ and $\kappa=0.02$ work well to bound the amount of computation for each additional edge: news items with few incoming edge propagate their change to few neighbors, and news items with a large number of incoming edges often do not change their value by a large amount as new edges are added. 

\begin{table}
\centering
    \begin{tabular}{r||r|r|r}
      & \multicolumn{2}{c|}{First-day} & Avg.\ 10-day\\
      & n.\ items & agreement        & agreement \\ \hline
      New Items   & 41,613 & 95.61\% & 95.66\% \\
      Tweeted Items & 62,002 & 96.53\%  & 96.57\% \\
      All Items   & 637,000 & 96.76\% & 96.59\%
    \end{tabular}
  \medskip
  \caption{Numbers and percentages of items with difference less than 10\% between on-line algorithm and daily fixpoint computation for the harmonic algorithm.}
  \label{table-online-classification}
\end{table}

\subsection{Importance of Reputation}

The methods and the results discussed above rely on the use of reputation systems, in order to associate a measure of trust to the sources and distributors and not just the content of a news article.
In this aspect, we differentiate from semantic analysis methods where a content analysis of the news articles can be used for detecting fake news.
A key concern in the problem of fake news is its evolutionary nature.
The methods to tamper or artificially generate a news sample are not limited by any metric, and can continue to evolve, further improving upon the semantic methods.
However, the reputation is organically learned from the interaction graph, and can always learn to identify sources spreading misinformation, even if the nature of the content changes over time.
Moreover, the semantic difference between a true and fake news report might not be significant in certain cases.
For example, reports that a certain elected official misbehaved can be both true and false, difficult to distinguish semantically, but looking at where the news comes from can be an important factor.
And it is a factor that we regularly use, whenever we use our trust in news sources to decide whether to believe them.
The methods presented in this paper preserves this trust in making the prediction between `fake' and `true' news.

\section{Discussion}

We believe that the challenge of building a reputation system for online news sources consists in creating an algorithm that is able to flag a large portion of fake news, while only minimally affecting mainstream news coming from publications with high journalistic standards and editorial controls. 

Of the algorithms considered in this paper, the harmonic algorithm is the one that performed the best in our tests, yielding correct classification of about 90\% for both fake and non-fake news. 
When using editorial edges, the mis-classification rate for mainstream reputable news sites was below 1\%.

In the harmonic algorithm, reputation propagates from news to users and then back to news items. 
In the traditional, batch implementation of this algorithm this feedback loop is iterated until it converges to a fixpoint.
We introduce an online version of the algorithm, which can classify URLs in real time by propagating information locally in the graph, and we show that its results deviate only in small measure from those of the offline, batch version.

Our tests also show that the users who spread a news item can be a more significant feature than the words appearing in the title or description of the news item itself, insofar as classifying the item is concerned. 
We believe this likely happens because on Twitter some users acts as indicators for particular types of news, systematically (re)tweeting a large portion of them. 
Indeed, we have reasons to suspect that some of the users having large influence (weights) in the logistic-regression based methods may be bots and this is something we intend to explore in future work.

Our results also highlight the ability of the algorithms to identify fake and misleading news sites that were not known to us in advance.
Indeed, some of the news sites in Table~\ref{table-news-sites} first came to the attention of the authors when looking for false positives.
This suggests an additional use of reputation systems for news: as a tool that helps human moderators and fact checkers find more sites that deserve inspection.


\subsection*{Acknowledgements.}
We thank Paul Walsh, CEO of Metacert, for making available to us the Metacert dataset of low-quality sites.

\bibliographystyle{alpha}
\bibliography{hoax}

\end{document}